# Surface Plasmon Driven Electric and Magnetic Resonators for Metamaterials

**Durdu Ö. Güney,**[1,2,*] **Thomas Koschny,**[1,3] **and Costas M. Soukoulis**[1,3]

[1]*Ames National Laboratory, USDOE and Department of Physics and Astronomy, Iowa State University, Ames, IA 50011*

[2]*Department of Electrical and Computer Engineering, Michigan Technological University, Houghton, MI 49931*

[3] *Institute of Electronic Structure and Laser, Foundation for Research and Technology Hellas (FORTH), and Department of Materials Science and Technology, University of Crete, 7110 Heraklion, Crete, Greece*

[*] dguney@mtu.edu

**Abstract:** Using interplay between surface plasmons and metamaterials, we propose a new technique for novel metamaterial designs. We show that surface plasmons existing on thin metal surfaces can be used to "drive" non-resonant structures in their vicinity to provide new types of electric and magnetic resonators. These resonators strictly adhere to surface plasmon dispersion of the host metal film. The operating frequency of the resultant metamaterials can be scaled to extremely high frequencies, otherwise not possible with conventional split-ring-resonator-based designs. Our approach opens new possibilities for theory and experiment in the interface of plasmonics and metamaterials to harvest many potential applications of both fields combined.

**Introduction**

Metamaterials and plasmonics are two rapidly growing areas of research motivated by their potential applications relevant to defense, security, communications, computing, energy, and health. Interplay between metamaterials and surface plasmons can lead to superior applications, such as ultra-high resolution imaging and high-precision lithography. Metamaterials are man-made, usually periodic structures, with subwavelength feature sizes. Given the capability to tailor the underlying effective electromagnetic constitutive parameters (ε and μ) of metamaterials almost arbitrarily [1], one can envision a wide range of interesting applications as varied as perfect lenses [2], invisibility cloaks [3], quantum levitation [4], compact antennas [5], and optical analogue simulators [6,7]. On the other hand, plasmonic waveguides, resonators, beam splitters, and interferometers can provide high density integration of optical circuits with broadband operation [8]. Surface plasmons, key to these miniaturized devices, are collective electronic charge oscillations residing usually at the metal-dielectric interfaces [9,10]. Surface plasmon polaritons (SPP) are the surface modes excited by the interaction of surface plasmons with photons. Their dispersion can be controlled by metamaterial surfaces [11,12]. Surface plasmons cannot interact directly with free space photons, due to the momentum mismatch. Common approaches to facilitate this interaction are attenuated total reflection [13-15], which makes use of evanescent coupling of total internally reflected photons to surface plasmons, gratings, or nanostructures introduced on or in the vicinity of metal surfaces [16,17].

Although reducing the geometric dimensions of conventional metamaterials can lead to higher operating frequency, it cannot be scaled to arbitrarily high operating frequencies [18-20]. This is not only due to difficulty in fabrication but also there is a physical limit which prohibits further scaling. At substantially high frequencies the effective mass of electrons becomes important. This breaks the linear scaling and, hence, the operating frequency saturates. Our plasmonic approach opens the possibility to scale the operating frequencies to visible frequencies and beyond. It is also worth mentioning that previously proposed excitations of surface plasmons using nanostructures rely on electric coupling of the incident field [16-17]. Here, in addition to electric excitation, we also show the magnetic excitation of surface plasmons on thin metal film surfaces for the first time. This can facilitate the designs of impedance matched metamaterials for highly demanded applications, in particular, high-efficiency solar energy concentration, high-resolution imaging [2,21], and high-precision lithography [22].

**Magnetic Resonator**

We start with the magnetic resonator and will describe the electric resonator later. To demonstrate this idea, we use the geometric design shown in Fig. 1a. This design consists of a thin film and *U*-shaped structures (*U*s) close to the surface of the film.

The surfaces of an infinitely thick metal film support degenerate surface modes. When the film thickness becomes finite, the degenerate mode splits and we observe two branches in the dispersion relation of the thin film (see Fig. 4 in Ref. 10). Both branches remain below the light line and, in the limit wavevector $k \rightarrow \infty$, converge to $f_p/\sqrt{2}$, where $f_p$ is the bulk plasma frequency for the metal film. The currents on the two surfaces of the film are anti-parallel for the upper branch and parallel for the lower branch—the anti-symmetric and symmetric modes, respectively. The small film thickness and the high dielectric constant of the background can be traded to pull down the lower branch to relatively lower operating frequencies. From the design and fabrication perspective, it is preferable to lower the operating frequency, but on the other hand, we would like to emphasize there is no theoretical limitation preventing the operating frequency from reaching extremely high frequencies, unlike the conventional split-ring-resonator based metamaterials.

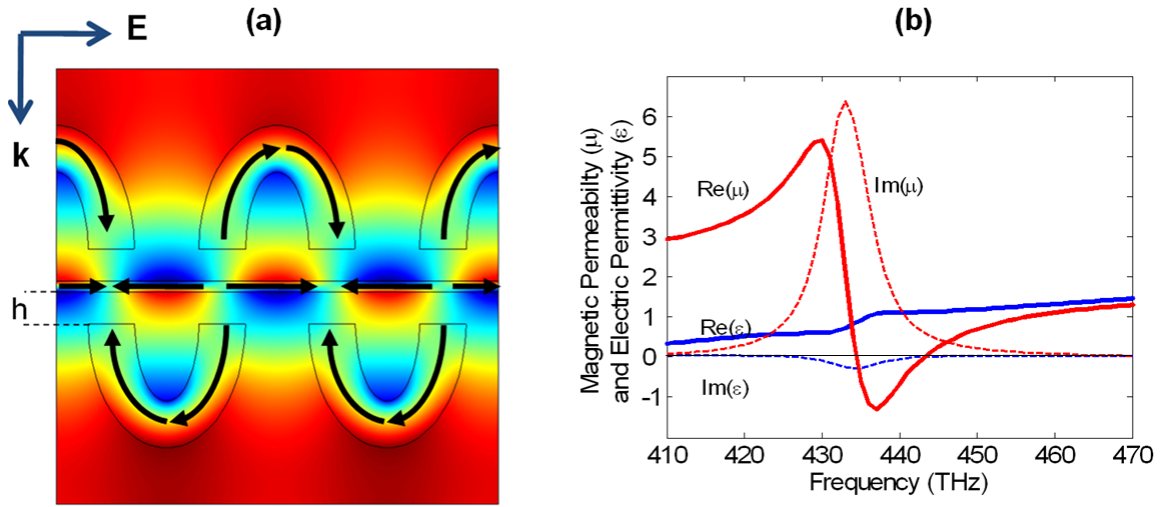

**Figure 1** (Color online) Surface plasmon coupled metallic *U*-shaped nanostructures as a magnetic resonator. (**a**) Surface plot shows the magnetic field distribution and the black arrows show the direction of currents under normal plane wave incidence. Incident field is polarized in horizontal direction to couple to the structure, which is translation invariant in the direction of magnetic field. The separation between the *U*s and the thin metal film is $h = 7.5$nm. The thickness of the thin film is 2.5nm. The thickness of the *U*s is 11nm and is uniform along their arms. The vertical distance between the tip of the *U*s and the film surface is 38nm. The length of the unit cell along the direction of propagation is 102.76nm. All metals are gold and described by the Drude model with $f_p = 2{,}175$THz and $f_c = 6.5$THz. The background is polyimide with $\varepsilon_r = 3.5$. Surface plasmon coupling to the *U*s lead to a net magnetic moment. (**b**)

Retrieved effective ε and μ using homogeneous effective medium approximation. μ is negative over a bandwidth of about 9THz.

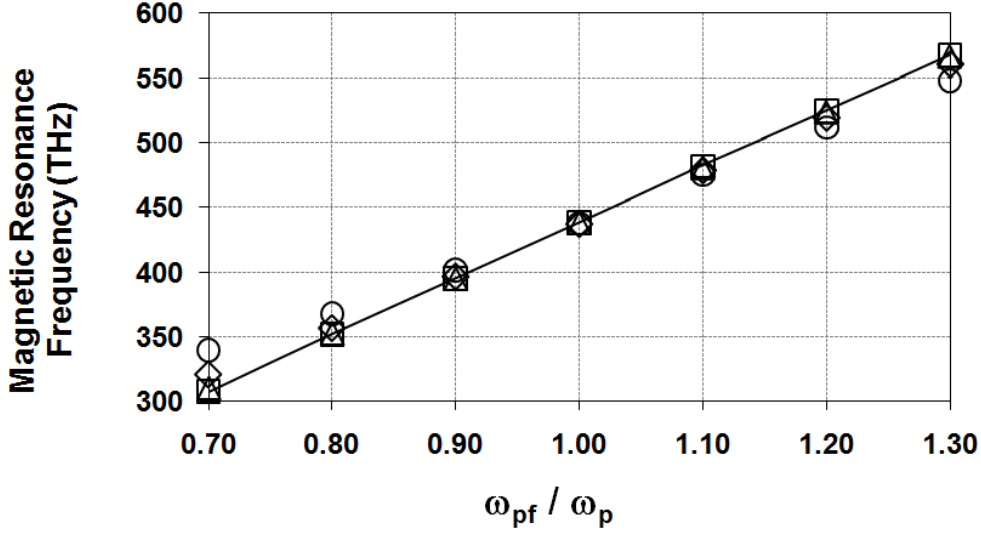

**Figure 2** (Color online) Dependence of magnetic resonance frequency on thin film permittivity parameterized by fictitious plasma frequency, $\omega_{pf}$. The solid line corresponds to the dispersion of the thin film alone. Different symbols correspond to different $h$ values, ○ ($h$ = 10nm), ◊ ($h$ = 15nm), Δ ($h$ = 25nm), and □ ($h$ = 30nm). $U$s disturb the surface mode more strongly at small $h$ values. Hence, the dispersion of the structure starts to deviate from the thin film dispersion. Conversely, as $h$ becomes large the coupling is reduced and the dispersion diagram converges to thin film dispersion.

For our numerical simulations we choose a gold film with thickness $t$ = 2.5nm in a polyimide background of relative permittivity, $\varepsilon_r$ = 3.5. We describe the gold film by the Drude model with the plasma and the collision frequencies taken as $f_p$ = 2,175THz and $f_c$ = 6.5THz [23]. Once we determine the corresponding dispersion relation of the thin film for above parameters, we choose a frequency point on the lower branch where we want our metamaterial to operate. The chosen frequency should be sufficiently far from the light line to avoid propagating diffraction orders and should simultaneously comply with the homogeneous effective medium approximation of the metamaterial with respect to the incident propagating waves. Under these considerations, we select the SPP frequency of $f_{SPP}$ = 438.3THz, which has a wavelength of $\lambda_{SPP}$ = 52.3nm. Indeed, the magnetic field profile for this eigenfrequency is very similar to the field distribution on the thin film shown in Fig. 1a. Because it is a symmetric mode, the currents on the two surfaces are parallel and are illustrated by single black arrows. We then place the $U$s close to the thin film. The thickness of the $U$s are 11nm and $h = 7.5$nm. The structure with only two unit cells in the horizontal direction and one unit cell in the vertical direction is shown. The $U$s are placed on both sides of the thin film in such a way they are $\lambda_{SPP}/2$ apart (i.e., center-to-center distance) in the horizontal direction. Other parameters are given in the caption of Fig. 1a. An incident plane wave with the given configuration, facilitated by the $U$s, excites the SPPs on the thin film, which otherwise cannot be excited by plane waves, due to the momentum mismatch. Once the SPP resonant mode is excited, it passively couples to the $U$s in its proximity and, hence, induces the circular currents which lead to a net magnetic moment.

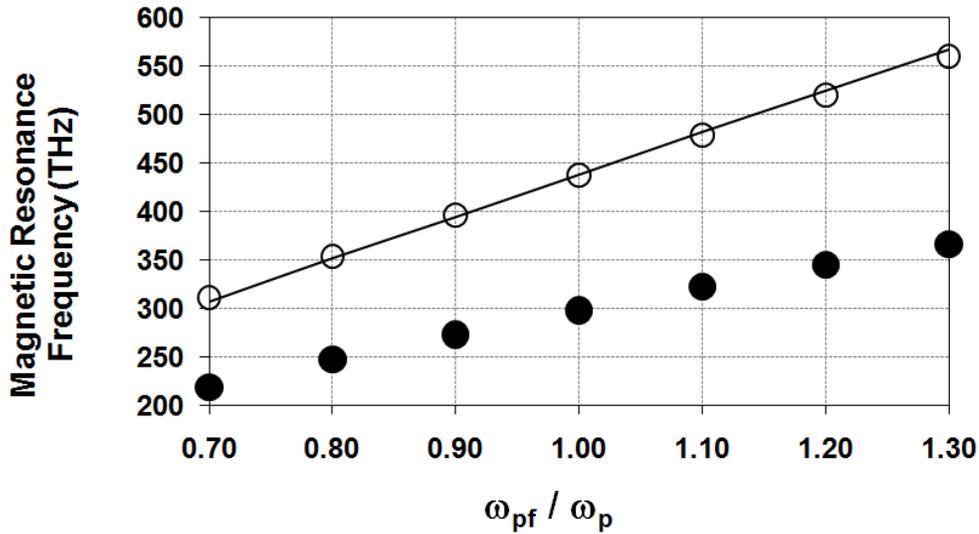

**Figure 3** (Color online) Magnetic resonance frequency versus metal permittivity parameterized by fictitious plasma frequency, $\omega_{pf}$. All metals are assumed to have the same permittivity for a given data point and $h = 7.5$nm. The solid line corresponds to the dispersion of the thin film alone. Magnetic resonance of the metamaterial and the self-resonances of the $U$s are shown by ○ and ● symbols, respectively.

We performed all simulations using the frequency domain finite element technique as implemented by commercially-available COMSOL Multiphysics software. One might expect cross-coupling between the electric and magnetic fields for the simulated structure in Fig. 1b, hence bianisotropy, since the structure has no inversion symmetry with respect to the propagation direction. However, the symmetry of the "corrugated layer" ensures vanishing electric moment in response to the normally incident magnetic field. So that under the given conditions the structure is actually not bianisotropic and we can use the isotropic retrieval procedure as described in Ref. [24] to retrieve effective ε and µ (shown Fig. 1b).

The region in Fig. 1b with negative imaginary values is due to the periodicity artifacts which occur if the system is not truly subwavelength and the resultant spatial dispersion is ignored in favor of purely frequency dependent effective medium parameters [25]. In this situation, negative imaginary parts arise in the effective parameters as consequence of this approximation. However, the negative imaginary parts observed in metamaterials do not violate requirements [26] of thermodynamics in homogeneous passive media. This is because metamaterials are never truly continuous media and are only defined in increments of a unit cell, allowing them more flexibility in the response functions without disturbing the passivity of the material. The retrieved effective permittivity and permeability accurately describe the scattering behavior of the actual metamaterial that is subject to such restrictions but should not be extrapolated to the generic response of an arbitrary continuum. In the limit of small lattice constant compared to wavelength inside the metamaterial those artifacts disappear.

The structure has magnetic resonance at 438THz with a very good agreement with the SPP resonance frequency. Effective µ is negative over about 9THz bandwidth reaching values below -1. The ratio of the resonant wavelength in free space to the structure size in the propagation direction is about seven (i.e., reasonably subwavelength structure size for the retrieval procedure). Because the SPP mode by itself has no net magnetic moment, the magnetic resonance here arises from the currents circulating on the *U*s. To verify that the obscured metamaterial resonance is indeed generated by the SPP resonance of the metal film and not by any intrinsic resonance of the *U*s, in Fig. 2, we assume generic metal films with different permittivity values and trace the resultant magnetic resonance frequency for different values of *h* (shown by different symbols). We keep the gold *U*s and describe the permittivity, $\varepsilon_m$, of the generic metal film by the Drude model as $\varepsilon_m = 1 - \omega_{pf}^2/(\omega^2 - j\omega\omega_c)$, where $\omega_c = 2\pi f_c$ and $\omega_{pf} = y 2\pi f_p$. $\omega_{pf}$ is a fictitious plasma frequency. As *y* changes between 0.7 and 1.3 in Fig. 2, the permittivity of the generic metal film also changes accordingly. For example, *y* = 1 corresponds to actual gold used in Fig. 1. Because we do not change the grating periodicity, for each $\omega_{pf}$ in Fig. 2 the resonance is observed at the frequency point where $\lambda_{SPP}$ matches the period of the surface grating. Although we assume the values of the plasma frequencies do not correspond to real metals, this method provides a means to understand observed behavior through such a fictitious dispersion relation.

Overall, the dispersion relation for different *h* values always closely follows the SPP dispersion, shown by the solid line. When the *U*s go further away from the thin film surface, they only negligibly disturb the surface mode, due to the weaker coupling, and, hence, the dispersion of the structure as a whole converges to the SPP dispersion. Simultaneously, the magnetic resonance also becomes weaker (not shown). Conversely, when the *U*s go closer to the thin film, the dispersion relation for the structure starts to deviate from the SPP dispersion, due to the strong disturbance on the surface modes (i.e., stronger coupling) and we obtain a strong magnetic resonance (see Fig. 1b). We also checked to determine if the *U*s have any resonance around the frequency ranges in Fig. 2 and found the self-resonances of the *U*s are always sufficiently far and decrease in frequency with increasing separation in the vertical direction. For completeness, in Fig. 3, we assume alternatively that the *U*s are made from the same fictitious metal as the thin film. The self-resonance of the *U*s, which is at approximately 300THz and independent of the fictitious plasma frequency in the case studied in Fig. 2, is shown (black filled circles) together with the resonance frequency of the metamaterial that has both *U*s and film made from the same fictitious metal (hollow circles) and the corresponding naked surface plasmon of the unperturbed film (solid line) in Fig. 3. As in this case the self-resonance of the *U*s also scales with the fictitious plasma frequency in the same direction as the resonance frequency of the surface plasmon at the wavevector corresponding to the given periodicity and both resonances remain well separated in frequency, the scaling of the metamaterial resonance now follows even more closely the dispersion of the naked surface plasmon of the undisturbed metal film. These results clearly show the magnetic resonance of the metamaterial structure is indeed driven by the SPPs and can be controlled by the underlying SPP dispersion of the thin film.

**Electric Resonator**

Next, we briefly discuss similar results for the electric resonator. Everything remains the same as in the case of magnetic resonator in Fig. 1, except we realign the gold *U*s as shown in Fig 4a. Such arrangement of the *U*s cancels the magnetic resonance as can be seen from the induced currents on the *U*s circulating in reverse directions. However, now we obtain a net electric dipole moment, due to the accumulating charges.

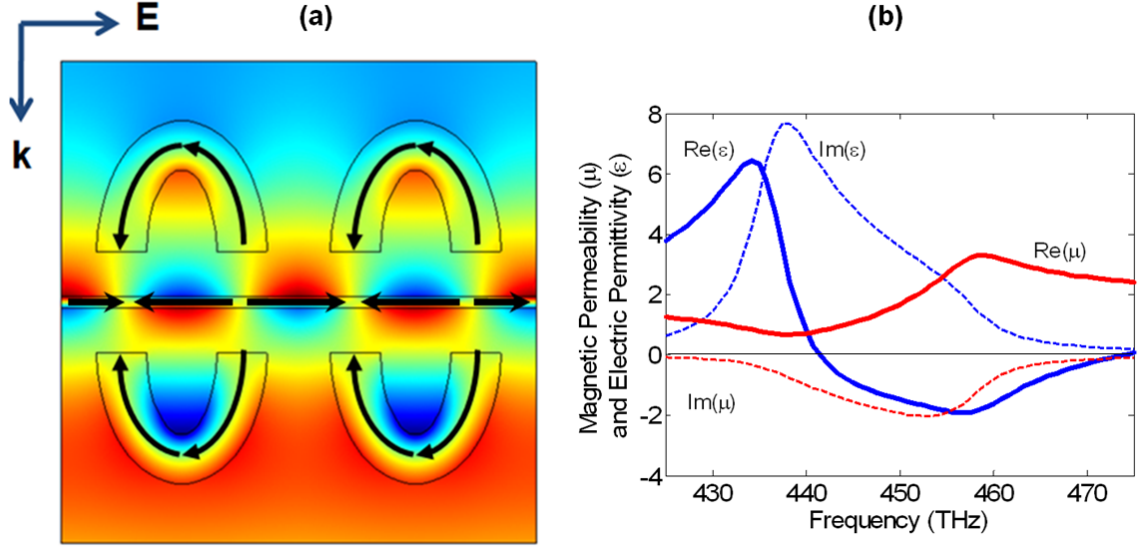

**Figure 4** (Color online) Surface plasmon coupled metallic *U*-shaped nanostructures as an electric resonator. (**a**) Surface plot shows the magnetic field distribution and the black arrows show the direction of currents under normal plane wave incidence. The separation between the *U*s and the thin metal film is $h = 10$nm. All other parameters are the same as in Fig. 1. Surface plasmon coupling to the *U*s results in a net electric moment. (**b**) Retrieved effective ε and μ using homogeneous effective medium approximation. ε is negative over a bandwidth of about 32THz.

The retrieved effective ε and μ are shown in Fig. 4b. Effective ε is negative over about 32THz bandwidth, more than three times broader compared with the bandwidth for negative magnetic response. Electric resonance frequency is at 437.5THz and again agrees very well with the SPP resonance frequency at 438.3THz. Due to the similar symmetry argument in the previous section, no cross-coupling terms exist in the permittivity and permeability tensors, so that we have used the isotropic retrieval procedure [24] to retrieve the corresponding effective parameters.

In Fig. 5, similar to the magnetic resonance explained above, we show the dependence of electric resonance frequency on the generic metal film permittivity. Further from the *U*s, the weaker the coupling and the electric resonances (not shown); and the dispersion relation converges to the SPP dispersion of the thin film (solid line). When the $U$s are close to the thin film surface, they start to destroy the SPP mode as seen by the deviation from the SPP dispersion. Similarly, these results show that with the correct orientation of the $U$s (see Fig. 4) we can also make a metamaterial electric resonator driven by the underlying SPP resonances of the thin metal film.

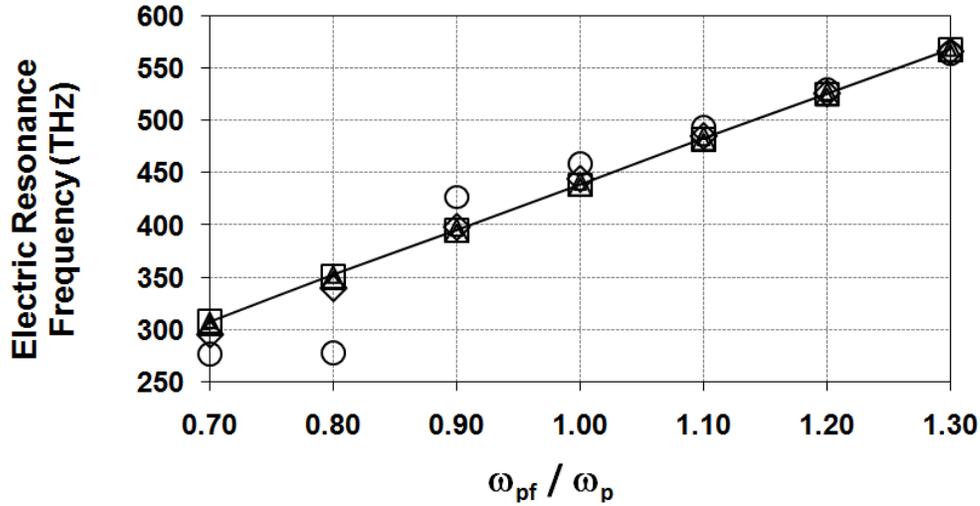

**Figure 5** (Color online) Dependence of electric resonance frequency on thin film permittivity parameterized by fictitious plasma frequency, $\omega_{pf}$. Different symbols correspond to different $h$ values—○ ($h$ = 10nm), ◊ ($h$ = 15nm), Δ ($h$ = 25nm), □ ($h$ = 30nm). $U$s disturb the surface mode more strongly at small $h$ values. Hence, the dispersion of the structure starts to deviate from the thin film dispersion. Conversely, as $h$ becomes large, the coupling is reduced and the dispersion diagram converges to thin film dispersion.

**Conclusions**

In conclusion, we proposed and numerically demonstrated a novel way of building metamaterials based on surface plasmon coupling to nearby nanostructures. The operating frequency of these metamaterials is given by the underlying surface modes, which provide guidance for the operating frequency through their dispersion. Thin metal films can provide extremely high operating frequencies. Paired structures such as fishnet structures or wire pairs can also provide high operating frequencies within the near-infrared or visible regions (around 700nm) [27]. However, the resonant response of these structures arises from localized geometric or particle plasmon resonances as in conventional split-ring-resonators. Therefore, the operating frequency of the paired structures is strongly dependent on the inherent geometric properties of the resonators and it is not obvious to what extent the frequency can be scaled up. In contrast, our proposed structure employs the coupling mechanism owing to the spatially extended surface plasmon modes of the thin metal film rather than the intrinsic particle plasmon modes of the nearby objects. For example, using the antisymmetric branch in the thin film dispersion diagram, operating frequencies can be scaled to even beyond visible spectrum with an accommodating fabrication technique (e.g., patterned planar metal-dielectric multilayer strips).

Experimental demonstration of the idea can be achieved at lower operating frequencies ranging from GHz to THz frequencies by using suitable materials. For example, polaritonic materials (e.g., SiC) and low-density plasmas (e.g., heavily doped semiconductors) can substitute for the thin metal film. Photonic crystal surfaces or corrugated surfaces, meta-surfaces in the sense of microstructured metallic patterns (e.g., nanorod or wires) can also be used as "drivers" for these metamaterial resonators to provide more flexibility in the dispersion engineering. Our approach opens new possibilities for theory and experiment at the interface of plasmonics and metamaterials to harvest the potential applications in both fields.

**Acknowledgement**

Work at Ames Laboratory was supported by the Department of Energy (Basic Energy Sciences) under Contract No.DE-AC02-07CH11358.This work was partially supported by AFOSR under MURI under Grant No.FA9550-06-1-0337.

**References**

[1] D. O. Guney, Th. Koschny, M. Kafesaki, and C. M. Soukoulis, Opt. Lett. **34**, 506 (2009).
[2] J. B. Pendry, Phys. Rev. Lett. **85**, 3966 (2000).

[3] D. Schurig, J. J. Mock, B. J. Justice, S. A. Cummer, J. B. Pendry, A. F. Starr, and D. R. Smith, Science **314**, 977 (2006).

[4] U. Leonhardt and T. G. Philbin, N. J. Phys. **9**, 254 (2007).

[5] I. Bulu, H. Caglayan, K. Aydin, and E. Ozbay, New J. Phys. **7**, 223 (2005).

[6] D. O. Guney and D. A. Meyer, Phys. Rev. A **79**, 063834 (2009).

[7] D. A. Genov, S. Zhang, and X. Zhang, Nature Phys. **5**, 687 (2009).

[8] S. I. Bozhevolnyi1, V. S. Volkov, E. Devaux, J.-Y. Laluet, T. W. Ebbesen, Nature **440**, 508 (2006).

[9] R. H. Ritchie, Phys. Rev. **106**, 874 (1957).

[10] E. N. Economou, Phys. Rev. **182**, 539 (1969).

[11] R. Ruppin, Phys. Lett. A **277**, 61 (2000).

[12] R. Ruppin, J. Phys. **13**, 1811 (2001).

[13] A. Otto, Z. Phys. **216**, 398 (1968).

[14] A. Otto, *Z. Phys.* **219**, 227 (1969).

[15] B. Wang, W. Dai, A. Fang, L. Zhang, G. Tuttle, Th. Koschny, and C. M. Soukoulis, Phys. Rev. B **74**, 195104 (2006).

[16] A. Ghoshal and P. G. Kik, J. Appl. Phys. **103**, 113111 (2008).

[17] A. Ghoshal and P. G. Kik, Appl. Phys. Lett. **94**, 251102 (2009).

[18] J. Zhou, Th. Koschny, M. Kafesaki, E. N. Economou, J. B. Pendry, and C. M. Soukoulis, Phys. Rev. Lett. **95**, 223902 (2005).

[19] M. W. Klein, C. Enkrich, M. Wegener, C. M. Soukoulis, and S. Linden, *Opt. Lett.* **31**, 1259 (2006).

[20] D. O. Guney, Th. Koschny, and C. M. Soukoulis, Phys. Rev. B **80**, 125129 (2009).

[21] Th. Koschny, R. Moussa, and C. M. Soukoulis, J. Opt. Soc. Am. B **23**, 485 (2006).

[22] T. Xu, Y. Zhao, J. Ma, C. Wang, J. Cui, C. Du, and X. Luo, Opt. Express **16**, 13579 (2008).

[23] M. A. Ordal, L. L. Long, R. J. Bell, S. E. Bell, R. R. Bell, Jr. R. W. Alexander and C. A. Ward, Appl. Opt. **22**, 1099 (1983).

[24] D. R. Smith, S. Schultz, P. Markos, and C. M. Soukoulis, Phys. Rev. B **65**, 195104 (2001).

[25] Th. Koschny, P. Markoš, E. N. Economou, D. R. Smith, D. C. Vier, and C. M. Soukoulis, Phys. Rev. B **71**, 245105 (2005).

[26] L. D. Landau and E. M. Lifshitz, Electrodynamics of continuous media, Pergamon Press, Oxford, 1960.

[27] V. M. Shalaev, Nature Photonics **1**, 41 (2007).